%% file: main.tex
\documentclass[preprint2]{aastex631}
\usepackage{CJK}

\newcommand{\Ebv}{\mbox{$E(B-V)$}}
\newcommand{\Egk}{\mbox{$E(G-K_\mathrm{S})$}}
\newcommand{\Ebprp}{\mbox{$E(G_\mathrm{BP}-G_\mathrm{RP})$}}
\newcommand{\Ehk}{\mbox{$E(H-K_\mathrm{S})$}}
\newcommand{\Av}{\mbox{$A_{V}$}}

\newcommand{\HI}{\mbox{$\mathrm{H}\,{\scriptstyle\mathrm{I}}$}}
\newcommand{\Wco}{\mbox{$W_\mathrm{CO}$}}
\newcommand{\Xco}{\mbox{$X_\mathrm{CO}$}}
\newcommand{\Htwo}{\mbox{$\mathrm{H}_2$}}
\newcommand{\HII}{\mbox{$\mathrm{H}\,{\scriptstyle\mathrm{II}}$}}

\newcommand{\NHI}{\mbox{$N_{\scriptsize\HI}$}}
\newcommand{\NHtwo}{\mbox{$N_{\mathrm{H}_2}$}}

\newcommand{\NH}{\mbox{$N_\mathrm{H}=\NHI+2\NHtwo$}}
\newcommand{\GDR}{\mbox{$\mathrm{GDR}$}}

\newcommand{\CO}{\mbox{CO}}
\newcommand{\COJ}{\mbox{\CO \;$J= 1 \rightarrow 0 $}}

\newcommand{\kms}{\mbox{\,km\,s$^{-1}$}}

\newcommand{\pscm}{\mbox{\,cm$^{-2}$}} 
\newcommand{\pscmpmag}{\mbox{\pscm\,mag$^{-1}$}} 

\newcommand{\M}{\mbox{$\mathcal{M}$}}
\newcommand{\D}{\mbox{$\mathcal{D}$}}

\received{2024 January 18}
\revised{2024 September 3}
\accepted{2024 September 3}

\submitjournal{AJ}

\shorttitle{Dust \& gas in molecular clouds}
\shortauthors{Li et al.}

\begin{document}
\begin{CJK*}{UTF8}{gbsn}

\title{The Correlation Between Dust and Gas Contents in Molecular Clouds}

  \correspondingauthor{Bing-Qiu Chen; Guang-Xing Li}
  \email{bchen@ynu.edu.cn; gxli@ynu.edu.cn}
  
  \author[0000-0002-4205-0933]{Rui-Zhi Li}
  \affiliation{Yunnan Observatories, 
    Chinese Academy of Sciences, 
    Kunming 650216, People's Republic of China}
  \affiliation{University of Chinese Academy of Sciences, 
    Beijing 101408, People's Republic of China}
  \affiliation{Department of Astronomy, 
    Yunnan University, 
    Kunming 650504, People's Republic of China}

  \author[0000-0003-2472-4903]{Bing-Qiu Chen}
  \affiliation{South-Western Institute for Astronomy Research, 
    Yunnan University, 
    Kunming 650504, People's Republic of China}

  \author[0000-0003-3144-1952]{Guang-Xing Li}
  \affiliation{South-Western Institute for Astronomy Research, 
    Yunnan University, 
    Kunming 650504, People's Republic of China}

  \author[0000-0001-9342-1485]{Bo-Ting Wang}
  \affiliation{Yunnan Observatories, 
    Chinese Academy of Sciences, 
    Kunming 650216, People's Republic of China}
  \affiliation{University of Chinese Academy of Sciences, 
    Beijing 101408, People's Republic of China}
  \affiliation{Department of Astronomy, 
    Yunnan University, 
    Kunming 650504, People's Republic of China}

  \author[0000-0001-5991-6118]{Hao-Ming Ren}
  \affiliation{Department of Astronomy, 
    Yunnan University, 
    Kunming 650504, People's Republic of China}

  \author[0009-0000-1247-8050]{Qi-Ning Guo}
  \affiliation{Department of Astronomy, 
    Yunnan University, 
    Kunming 650504, People's Republic of China}


\begin{abstract}
Molecular clouds are regions of dense gas and dust in space where new stars and planets are born. There is a strong correlation between the distribution of dust and molecular gas in molecular clouds. The present work focuses on the three-dimensional morphological comparisons between dust and gas within 567 molecular clouds identified in a previously published catalog. We confirm a sample of 112 molecular clouds, where the cloud morphology based on CO observations and dust observations displays good overall consistency. There are up to 334 molecular clouds whose dust distribution might be related to the distribution of gas. We are unable to find gas structures that correlate with the shape of the dust distribution in 24 molecular clouds. For the 112 molecular clouds where the dust distribution correlates very well with the distribution of gas, we use CO observational data to measure the physical properties of these molecular clouds and compare them with the results derived from dust, exploring the correlation between gas and dust in the molecular clouds. We found that the gas and dust in the molecular clouds have a fairly good linear relationship, with a gas-to-dust ratio of $\mathrm{GDR}=(2.80_{-0.34}^{+0.37})\times10^{21}\mathrm{\,cm^{-2}\,mag^{-1}}$. The ratio varies considerably among different molecular clouds. We measured the scale height of dust-CO clouds exhibiting strong correlations, finding $h_{Z} = 43.3_{-3.5}^{+4.0}\mathrm{\,pc}$.
\end{abstract}



\section{Introduction} \label{sec:intro}

Molecular clouds (MCs), which are the sites of star formation, exhibit significantly higher extinction than the ambient diffuse interstellar medium (ISM) owing to their dense gas and dust contents. Understanding the interplay between gas and dust within MCs is pivotal for unraveling the intricacies of galactic evolution, the stellar initial mass function, and the star formation process as a whole \citep{2003ARA&A..41..241D, 2015ARAA..53..583H,2017ASSL..442.....M,2024FPP....1100059F}. A robust correlation between molecular gas and dust has been observed in these clouds, highlighting a deeply interwoven relationship. Dust is instrumental in star formation, serving as a shield for molecules against ultraviolet (UV) photodissociation, facilitating the cooling of gas, and enabling fragmentation, which leads to star birth \citep{2009ApJ...699..850K, 2013EP&S...65..213A, 2018MNRAS.474.4672L}. At the peripheries of MCs, the gas primarily consists of atomic hydrogen (\(\HI\)). In MCs with lower metallicity, UV photons from massive stars penetrate deeper, photodissociating CO and ionizing carbon to form \(\mathrm{C}^+\) \citep{2002ApJ...579..270P}. This results in an extended \(\mathrm{C}^+\)-emitting envelope surrounding a more compact CO core, while molecular hydrogen (\(\Htwo\)) is photodissociated by absorbing Lyman-Werner band photons \citep{2024A&A...688A.171K}. In denser regions of MCs, \(\Htwo\) can become optically thick, enabling it to self-shield or be shielded by dust against photodissociation, though CO remains photodissociated. Consequently, CO forms at higher opacities within the photodissociation regions (PDRs) than \(\Htwo\) \citep{2003ARA&A..41..241D}. A portion of molecular hydrogen thus exists beyond the CO-emitting region, commonly referred to as CO-dark \(\Htwo\) gas \citep{2002ApJ...579..270P,2005Sci...307.1292G, 2010ApJ...716.1191W}. Far-infrared (FIR) emission, corrected for dust temperature, serves as an excellent tracer for molecular hydrogen, as demonstrated by \citet{1998ApJ...500..525S}. Nonetheless, dust observations do not yield insights into the kinematics or dynamics of MCs \citep{2022ApJ...931....9L}. Comprehensive observations of both gas and dust are needed to delineate MCs based on dust and to ascertain their physical attributes. 

Prior investigations have predominantly targeted high Galactic latitudes \citep[e.g.,][]{2014ApJ...780...10L,2014ApJ...783...17L,2017ASSL..442.....M,2024ApJ...961..204S,2024A&A...688A.171K}, where the line-of-sight generally intersects a single MC, simplifying the quantification of dust and gas amounts. For accurate quantification, optimal sight lines should exhibit minimal CO presence and be situated at high ecliptic latitudes, where zodiacal light minimally contributes to the total FIR emission \citep{2016ApJ...821...78S}. At high Galactic latitudes, FIR emission strongly correlates with \(\HI\) observations until an excess in FIR emission indicates the formation of \(\Htwo\). This may explain why the ratio of neutral hydrogen column density to reddening, \(\NHI/\Ebv\), remains relatively constant in high Galactic latitude regions devoid of significant CO \citep{2014ApJ...780...10L, 2017ApJ...846...38L, 2020A&A...639A..26K}. Both dust emission and reddening are fundamentally governed by the optical properties of dust grains \citep{2016ApJ...821...78S}. The FIR emission from dust (\(I_\mathrm{FIR}\)) and the reddening (\(\Ebv\)) are strongly correlated with the hydrogen column density (\(N_\mathrm{H}\)) and are expected to scale linearly with it \citep{2014ApJ...780...10L, 2014ApJ...783...17L, 2020A&A...639A..26K}. However, at low Galactic latitudes, even simple measurements, such as FIR optical depth, pose significant challenges \citep{2014ApJ...789...15S, 2016ApJ...821...78S}. Within the Galactic plane ($|b| \leq 10^\circ$), the blending of emissions from multiple overlapping MCs complicates the isolation of signals from individual clouds \citep{2018MNRAS.474.4672L, 2020MNRAS.493..351C}. Consequently, accurate measurements of MCs within the Galactic plane have been elusive.

The abundance and relative optical thinness of dust, combined with its propensity to mix well with \(\mathrm{H_2}\), make it a powerful tracer of molecular hydrogen, surpassing CO in this role and thus providing a reliable means for probing MCs \citep{2009ApJ...692...91G, 2017MNRAS.472.3924C, 2022ApJ...931....9L}. The advent of extensive multi-band photometry for stars, coupled with precise stellar distance estimates from the \textit{Gaia} mission \citep{2018A&A...616A...1G}, has recently facilitated the accurate determination of distances and extinctions for individual MCs through the extinction breakpoint method \citep[e.g.,][]{2014ApJ...783..114G,2014ApJ...786...29S, 2019A&A...625A.135L, 2019MNRAS.488.3129Y, 2019ApJ...879..125Z, 2020A&A...633A..51Z, 2020MNRAS.496.4637C, 2020MNRAS.493..351C, 2021ApJS..256...46S, 2022MNRAS.511.2302G,2024A&A...685A..39M}. The method for estimating distances to dust clouds involves analyzing relative changes in stellar reddening caused by dust within the clouds. This technique infers the reddening and distance of stars using photometric data and parallax measurements. By modeling the cloud as a simple dust screen and identifying where a significant ``break'' in reddening occurs between unreddened foreground stars and reddened background stars, one can estimate the cloud's reddening and distance \citep{2019ApJ...879..125Z, 2020A&A...633A..51Z}. This method accounts for uncertainties in reddening and distance estimates, which are influenced by factors such as the spatial density of targets and uncertainties in photometric and parallax measurements, often using probabilistic models \citep{2019A&A...625A.135L}. The extinction breakpoint method offers significant advantages over traditional dust measurement techniques, which face challenges such as substantial variations in dust optical depth and spectral energy distributions, as well as difficulties in achieving optimal measurement conditions within the Galactic plane. By directly inferring distances to MCs through analyzing changes in stellar reddening between foreground and background stars, and employing probabilistic models to account for uncertainties, the extinction breakpoint method provides more robust and reliable estimates of reddening and distance, especially for MCs within the Galactic plane. It is less sensitive to multiple dust layers or intervening light sources, making it effective under various observational conditions. These advancements enable the delineation of MCs within the Galactic plane based on dust morphology and, when combined with gas data, allow for comprehensive measurement of their physical properties. Furthermore, this methodology allows for rigorous validation of dust tracer findings through direct comparison with real gas tracers, providing new insights into the radial velocity structure of the identified excess reddening clouds.

Several individual case studies have analyzed the dust-gas relationship in noted MCs by employing dust and CO observations, such as in the Pipe nebula \citep[e.g.,][]{2006A&A...454..781L}, Perseus \citep[e.g.,][]{2008ApJ...679..481P, 2014ApJ...784...80L}, Taurus \citep[e.g.,][]{2010ApJ...721..686P}, Orion \citep[e.g.,][]{2013MNRAS.431.1296R}, and California \citep[e.g.,][]{2015ApJ...805...58K, 2021ApJ...908...76L}. However, such direct observations of dust and gas are limited to a small subset of local MCs. Utilizing extinction maps, CO, and $\HI$ observations, \citet{2015MNRAS.448.2187C} probed the association between extinction and the emissions from $\HI$ and CO for eight giant MCs directed toward the Galactic anti-center. With data from the \textit{Planck} mission, \citet{2018MNRAS.474.4672L} quantified the relationship between CO's integrated intensity and visual extinction at parsec scales in 24 local MCs. A systematic examination of the dust and CO emission correlation across 12 local MCs was conducted by \citet{2022ApJ...931....9L}. These studies, nonetheless, focus on relatively small, local samples.

In this study, we focus on the three-dimensional (3D) morphological comparisons between dust and gas in 567 MCs identified by \citet{2020MNRAS.493..351C}. We further attempt to provide the radial velocity ranges of correlated dust-CO clouds within the Galactic plane, which may help resolve superimposed clouds along a single line-of-sight. The structure of this paper is as follows. Section \ref{sec:data} introduces the catalog of 567 MCs within the Galactic plane ($|b| < 10^\circ$), as well as the dust and gas data employed. Section \ref{sec:method} outlines our methodology for the morphological identification of MCs using the aforementioned data and details the derivation of the physical properties for each cloud. Our findings and their implications are discussed in Section \ref{sec:results and discussions}. Finally, we provide a summary in Section \ref{sec:summary}.

\section{Data} \label{sec:data}

\subsection{Catalog of 567 MCs within the Galactic Plane}\label{sec:MC_Catalog}

For this study, we use the catalog of 567 MCs within the Galactic plane identified by \citet{2020MNRAS.493..351C}. These MCs were detected using a hierarchical structure identification method implemented in the Python package \texttt{astrodendro}\footnote{\url{https://dendrograms.readthedocs.io}}, which was applied to 3D dust reddening maps created by \citet{2019MNRAS.483.4277C}.

\citet{2019MNRAS.483.4277C} utilized photometric data from over 56 million stars at low Galactic latitudes ($|b|\leq10^{\circ}$). The dataset combined optical photometry (in the $G$, $G_\mathrm{BP}$, and $G_\mathrm{RP}$ bands) from \textit{Gaia} DR2 \citep{2018A&A...616A...1G}, near-infrared photometry (in the $J$, $H$, and $K_\mathrm{S}$ bands) from the Two Micron All Sky Survey \citep[2MASS;][]{2006AJ....131.1163S}, and the $W1$ band from the Wide-Field Infrared Survey Explorer \citep[WISE;][]{2010AJ....140.1868W,2014ApJ...783..122K}. This combination helps to resolve the degeneracy between intrinsic colors and extinction for individual stars. The color excesses $\Egk$, $\Ebprp$, and $\Ehk$ for each star were determined using a Random Forest regression algorithm, trained on a comprehensive spectroscopic dataset of over 3 million stars. The typical uncertainty in these color excess values is around 0.07 mag for $\Ebv$ \citep[see Section 5.2 of][]{2019MNRAS.483.4277C}. By incorporating distance estimates from \citet{2018AJ....156...58B}, who inferred distances from \textit{Gaia} DR2 parallaxes for 1.33 billion stars using a Bayesian framework, \citet{2019MNRAS.483.4277C} constructed detailed 3D dust reddening maps across the Galactic plane. Using these maps, \citet{2020MNRAS.493..351C} identified 567 MCs via the hierarchical structure identification approach and determined their distances using the extinction breakpoint method, achieving uncertainties below 5\% (for a discussion on distance uncertainty, see Section 5.1 of \citet{2020MNRAS.493..351C}). The morphological details of these clouds are provided as downloadable contour maps\footnote{\dataset[Chen+2020\_allcloud.pdf]{\doi{10.12149/101367}}}, and a summary table\footnote{\dataset[Chen+2020\_table1.dat]{\doi{10.12149/101367}}} of their physical properties is available at doi:\dataset[10.12149/101367]{\doi{10.12149/101367}}. For the criteria used by \citet{2020MNRAS.493..351C} to identify MCs from 3D dust maps, they employed a boundary threshold of 0.05\,mag\,kpc$^{-1}$ for $\Egk$ and a significance threshold of 0.06\,mag\,kpc$^{-1}$ for $\Egk$. Furthermore, \citet{2020MNRAS.493..351C} calculated the $A_{V}^\mathrm{min}$ values, which define the polygons used in this work for visual identification of ${^{12}}$CO counterparts. These values are listed in the ``avmin'' entry in the last column of the summary table. This catalog provides an exemplary sample for investigating dust and gas within the Galactic MCs.

\subsection{$\COJ$ Data}
The $\COJ$ emission line data at 115 GHz, covering Galactic latitudes within $|b| \lesssim 30^{\circ}$, are presented by \citet{2001ApJ...547..792D}. Compiled from 37 individual surveys, these observations aim to maintain consistent rms noise levels by dynamically adjusting the integration time for each scan based on the real-time system temperature, assessed through 1-second calibrations \citep[see Section 2.1 of][]{2001ApJ...547..792D}. To accommodate the extensive velocity span of CO emission within the Galactic plane, flat spectral baselines are ensured through frequent position or frequency switching techniques. The surveys achieved an average angular resolution of $\theta_\mathrm{FWHM} \approx 8\arcmin.5$, with a local standard of rest (LSR) velocity coverage up to $|v_\mathrm{LSR}| < 332\kms$ and a velocity resolution of $\Delta v = 1.3\kms$. These composite maps provide a detailed representation of the molecular gas structure within the Galaxy and are instrumental for studying the interrelation of gas, dust, Population I objects, and young stellar objects within MCs. The complete dataset is accessible online, and for our purposes we utilize the position-position-velocity (PPV) data cube at latitudes $|b|\leq10^{\circ}$. 

\section{Method} \label{sec:method}

Our study is designed to conduct a 3D comparison of morphological features between dust and gas within Galactic MCs. We utilize the CO emission data from \citet{2001ApJ...547..792D} to coincide with MCs already identified via 3D dust contours. For compatibility, we resample the CO data cube to align with the 3D dust reddening map provided by \citet{2019MNRAS.483.4277C} using the Python package \texttt{reproject} (v0.7.1\footnote{\url{https://reproject.readthedocs.io}}). The resampled data spans Galactic latitudes within $|b|\leq10^\circ$ and Galactic longitudes ranging from $0^\circ<l<360^\circ$.

\begin{figure*}[htbp]
    \centering
    \includegraphics[scale=.7]{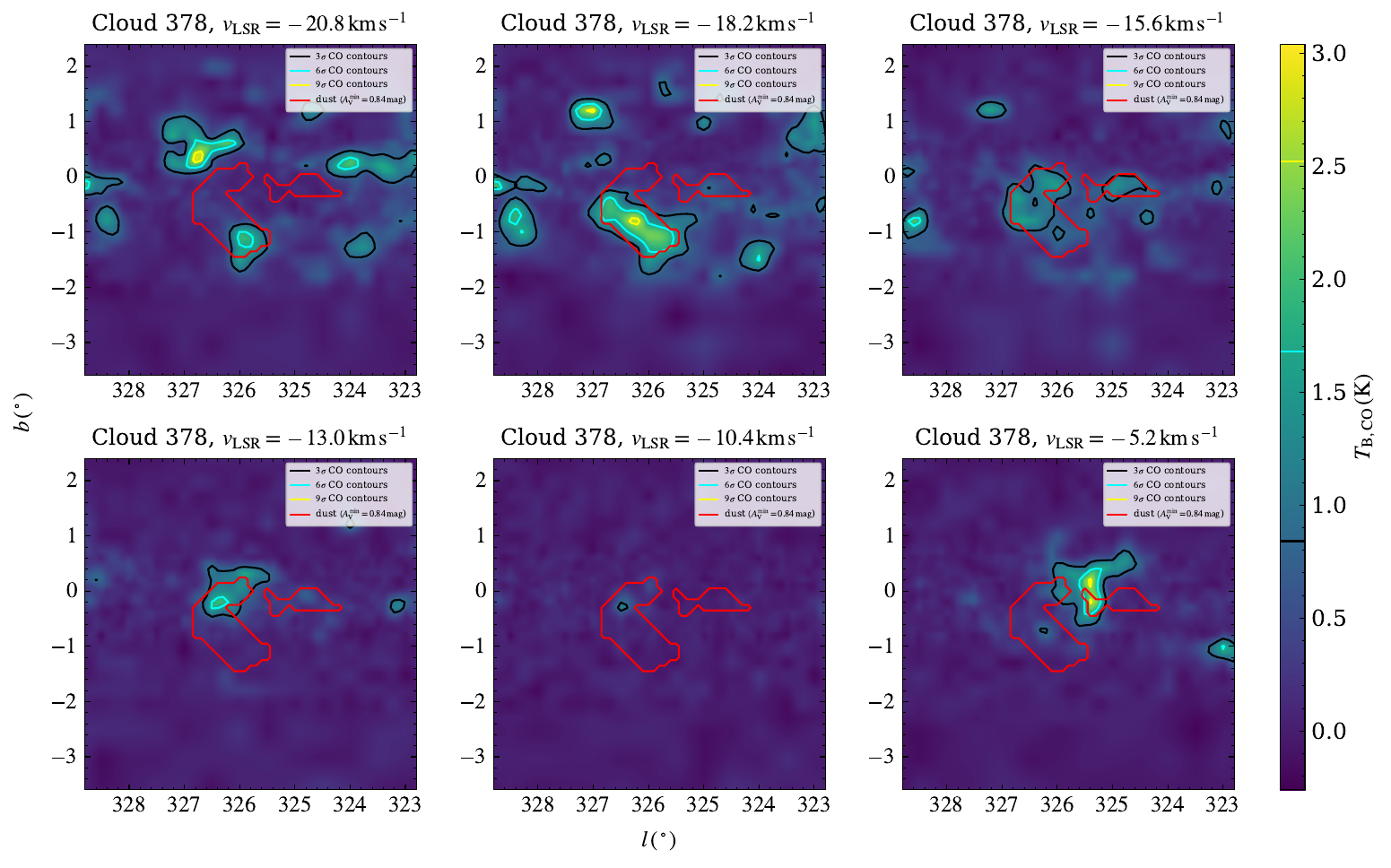}
    \caption{CO brightness temperature maps for Cloud 378 at various $v_\mathrm{LSR}$. The black, cyan, and yellow polygons denote CO emission significance at $3\sigma$, $6\sigma$, and $9\sigma$, respectively. The lines in the color bar indicate the CO emission at different levels. The red polygon highlights the dust contour boundary of the cloud as identified by \citet{2020MNRAS.493..351C}. The method for calculating the dust contour threshold, $A_{V}^\mathrm{min}$, is described in Section~\ref{sec:MC_Catalog}.}
    \label{fig:each_channel_MC_378}
\end{figure*}

\begin{figure*}[htbp]
    \centering
    \includegraphics[scale=.7]{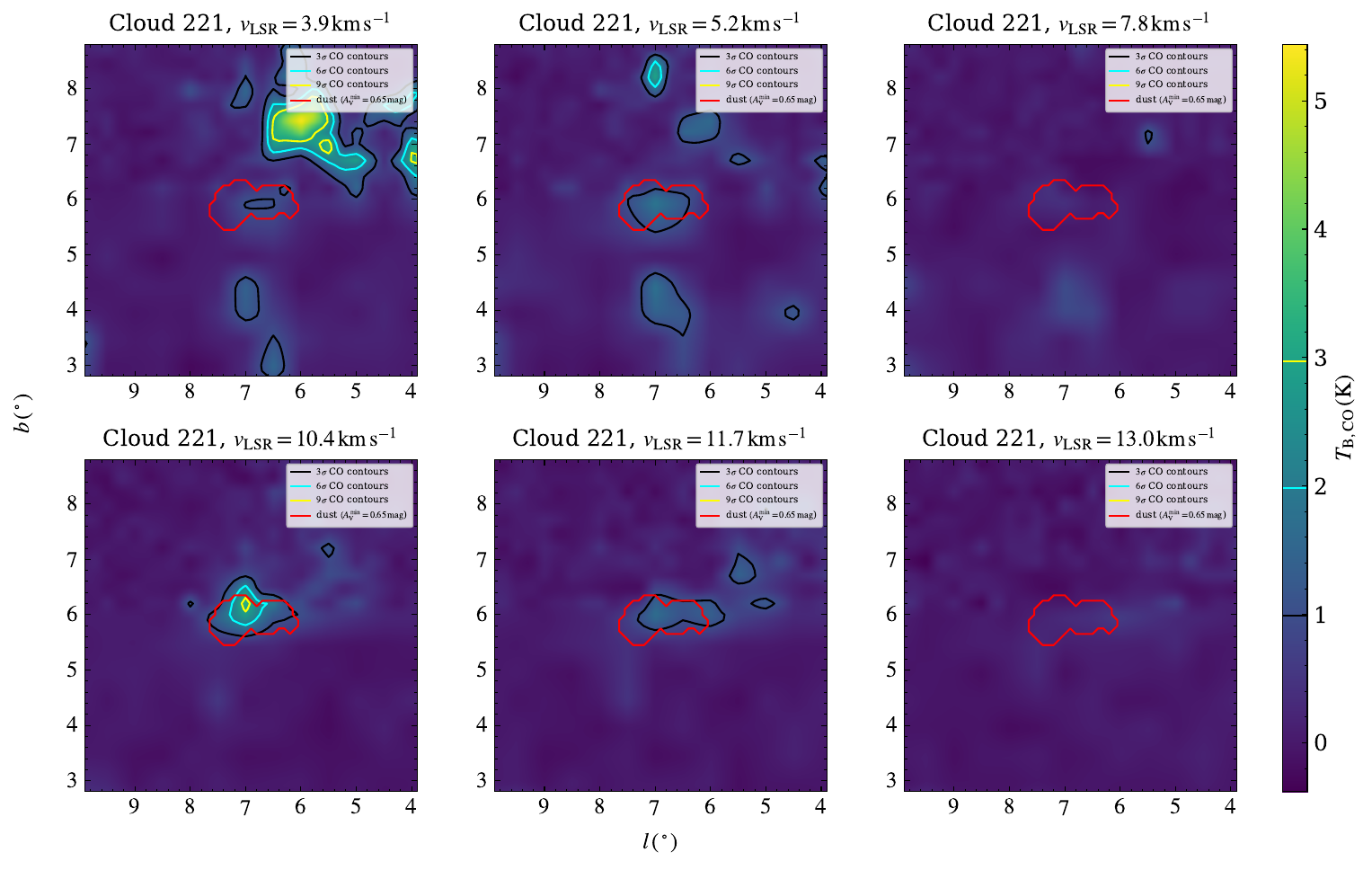}
    \caption{CO brightness temperature maps for Cloud 221, analogous to Figure~\ref{fig:each_channel_MC_378}.}
    \label{fig:each_channel_MC_221}
\end{figure*}

\begin{figure*}[htbp]
    \centering
    \includegraphics[scale=.7]{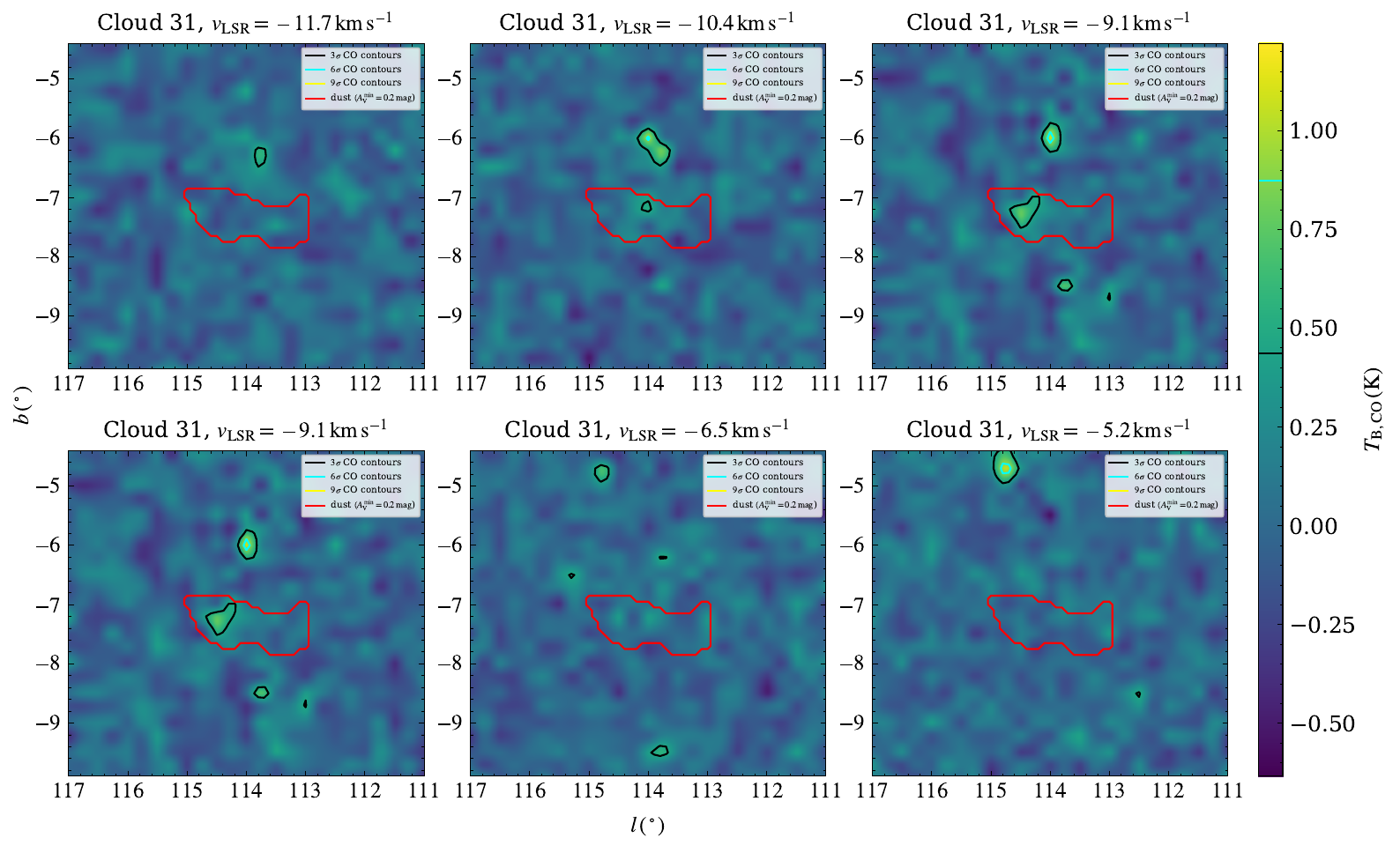}
    \caption{CO brightness temperature maps for Cloud 31, following the format of Figure~\ref{fig:each_channel_MC_378}.}
    \label{fig:each_channel_MC_31}
\end{figure*}

The initial phase of our analysis involves determining the LSR velocity ($v_\mathrm{LSR}$) range for each MC. This is accomplished by a visual inspection that compares the CO brightness temperature across velocity channels against the spatial boundaries of MCs as cataloged by \citet{2020MNRAS.493..351C}. We illustrate this process with case studies for Clouds 378, 221, and 31, as shown in Figures \ref{fig:each_channel_MC_378}, \ref{fig:each_channel_MC_221}, and \ref{fig:each_channel_MC_31}, respectively. 

To assess the similarity between the dust and CO contours, we utilized the \texttt{cv2.matchShapes} function from the \texttt{OpenCV-Python} (v4.10.0\footnote{\url{https://docs.opencv.org/4.10.0/index.html}}) library, a tool extensively employed in computer vision. This function relies on Hu moments, which are invariant under transformations such as translation, rotation, and scaling. By comparing the Hu moments of the input contours, the \texttt{cv2.matchShapes} function calculates a similarity score. Among the available methods for measuring differences, \texttt{cv2.CONTOURS\_MATCH\_I1} is frequently used. This method computes the sum of the absolute differences between the reciprocals of the logarithms of the Hu moments of the two contours. Once the contours are identified, the \texttt{cv2.matchShapes} function is deployed to determine a similarity score, with lower scores indicating greater similarity. This method effectively leverages the geometric invariance of Hu moments, making it a robust tool for shape matching and recognition across various image analysis applications. However, the dependency of the similarity score on the threshold level selection for the CO contour introduces variability, complicating the provision of a fixed quantitative measure of similarity.

The morphological identification of gas and dust contours in MCs involves some subjectivity due to the intricate geometry and dynamics of the clouds, necessitating a case-by-case comparison. Consequently, we evaluate the correlation between dust and CO contours, primarily through visual inspection rather than relying on a definitive similarity score threshold. In the cases of Clouds 378 and 221 (Figures~\ref{fig:each_channel_MC_378} and \ref{fig:each_channel_MC_221}), we observe good correlations between the dust contours and the CO emission maps at specific $v_\mathrm{LSR}$ ranges. Conversely, no similar gas structures corresponding to the dust distribution are apparent in Cloud 31 (Figure~\ref{fig:each_channel_MC_31}).

For each MC under investigation, we visually determine the $v_\mathrm{LSR}$ range within which the gas is correlated with the dust. Subsequently, the CO PPV data cube is integrated over this velocity range to compute the cloud's total CO intensity, $W_\mathrm{CO}$, as follows:
\begin{equation}
    W_\mathrm{CO} = \int T_\mathrm{B,CO}(v_\mathrm{LSR}) \, \mathrm{d}v_\mathrm{LSR},
\end{equation}
where $T_\mathrm{B,CO}(v_\mathrm{LSR})$ signifies the brightness temperature of the CO emission line at velocity $v_\mathrm{LSR}$.

\begin{figure*}[htbp]
\centering
    \gridline{\fig{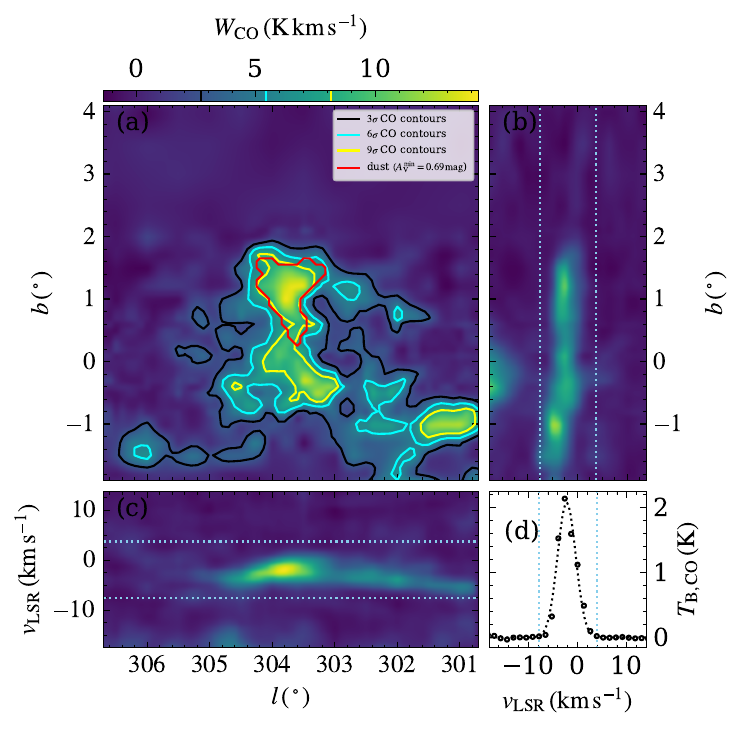}{0.33\textwidth}{(1) Cloud 89}
              \fig{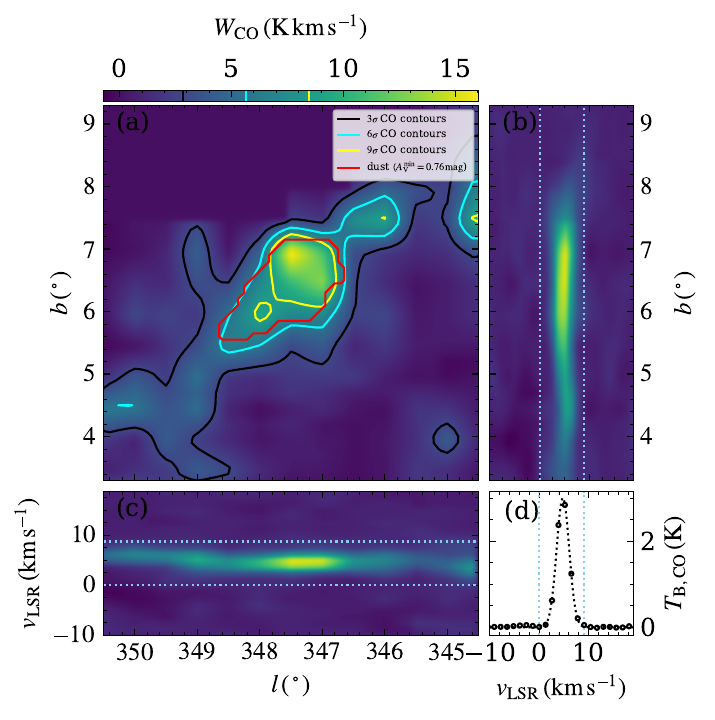}{0.33\textwidth}{(2) Cloud 137}
              \fig{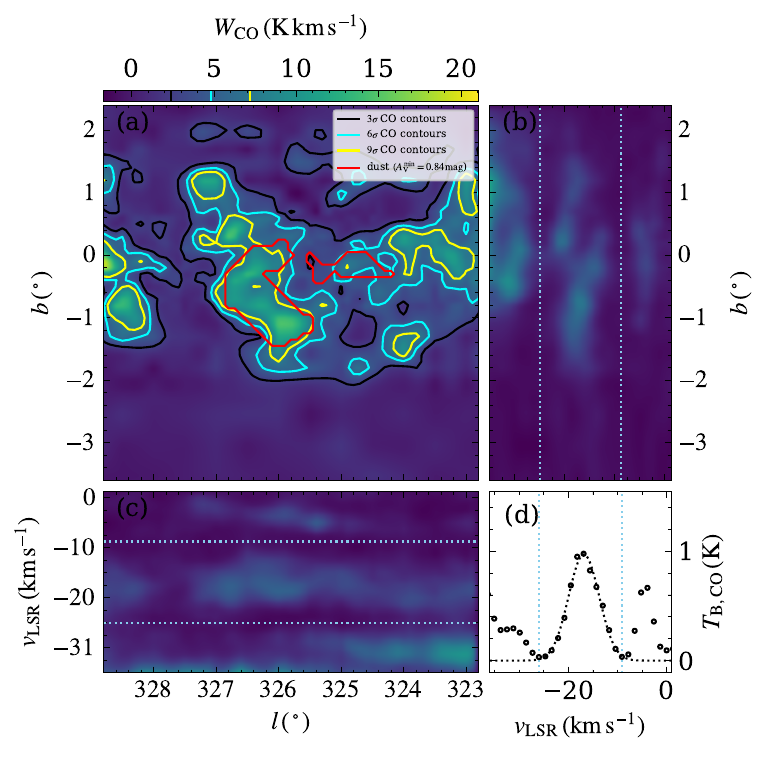}{0.33\textwidth}{(3) Cloud 378}}
    \caption{Morphological identification maps of selected strongly correlated dust-CO clouds. For each cloud, panel (a) displays the velocity-integrated CO intensity map. The black, cyan, and yellow polygons represent the velocity-integrated CO intensity levels at $3\sigma$, $6\sigma$, and $9\sigma$, respectively. The lines in the color bar indicate the CO integrated intensity at different levels. The red polygon highlights the dust contour boundary. Panel (b) represents the CO intensity integrated along the Galactic longitude $l$, panel (c) along the Galactic latitude $b$, and panel (d) shows the averaged CO line velocity-temperature profile for the respective MC, with the sky blue-dotted line indicating the $v_\mathrm{LSR}$ range.}
    \label{fig:dust-CO well correlated}
\end{figure*}

Using the velocity-integrated CO intensity maps, we then proceed to identify MCs that show any correlation between dust and CO emissions. To define a cloud as a ``strongly correlated dust-CO cloud,'' the CO emission line integrated intensity map must not only correspond closely with the dust contour morphology but also exhibit a single-peaked Gaussian velocity-temperature profile, indicating no overlap with clouds at other velocities. Figure~\ref{fig:dust-CO well correlated} illustrates examples of such MCs where dust and CO distributions are highly correlated. In the right-hand panel of Figure~\ref{fig:dust-CO well correlated}, which illustrates Cloud 378, the superposition of multiple CO peaks along a single line-of-sight is apparent. Velocity channel maps, shown in Figure~\ref{fig:each_channel_MC_378}, reveal that the CO peak at $v_\mathrm{LSR} \sim -18.2 \ \mathrm{km \ s^{-1}}$ aligns well with the dust contour. However, CO peaks at more distant ($v_\mathrm{LSR} \sim -30 \ \mathrm{km \ s^{-1}}$) and closer ($v_\mathrm{LSR} \sim -5.2 \ \mathrm{km \ s^{-1}}$) velocities do not match the dust contours. To emphasize this discrepancy, the CO brightness temperature map of the component at $v_\mathrm{LSR} \sim -5.2 \ \mathrm{km \ s^{-1}}$ is presented in the bottom right-hand panel of Figure~\ref{fig:each_channel_MC_378}. This observation demonstrates that superimposed clouds along a single line-of-sight can be resolved, which is a primary objective of this study. Consequently, Cloud 378 is classified as a strongly correlated dust-CO cloud.

\begin{figure*}[htbp]
    \centering
    \gridline{\fig{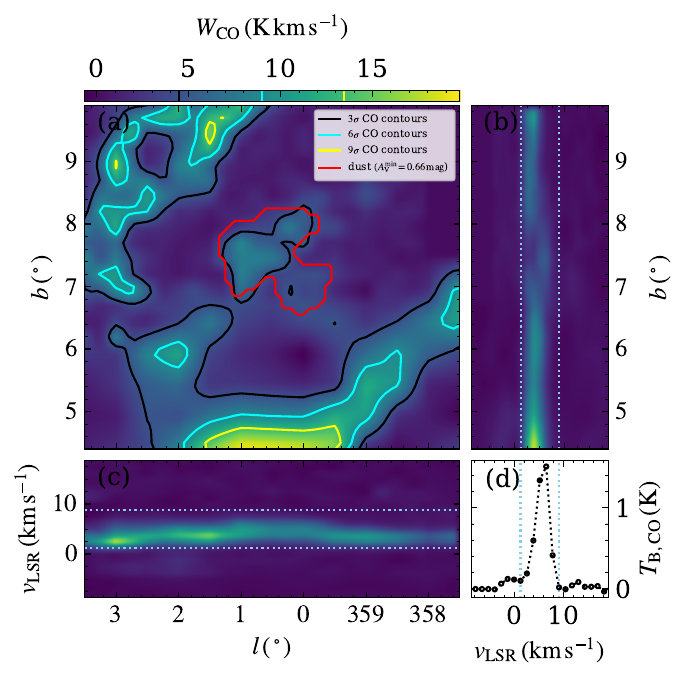}{0.33\textwidth}{(1) Cloud 158}
              \fig{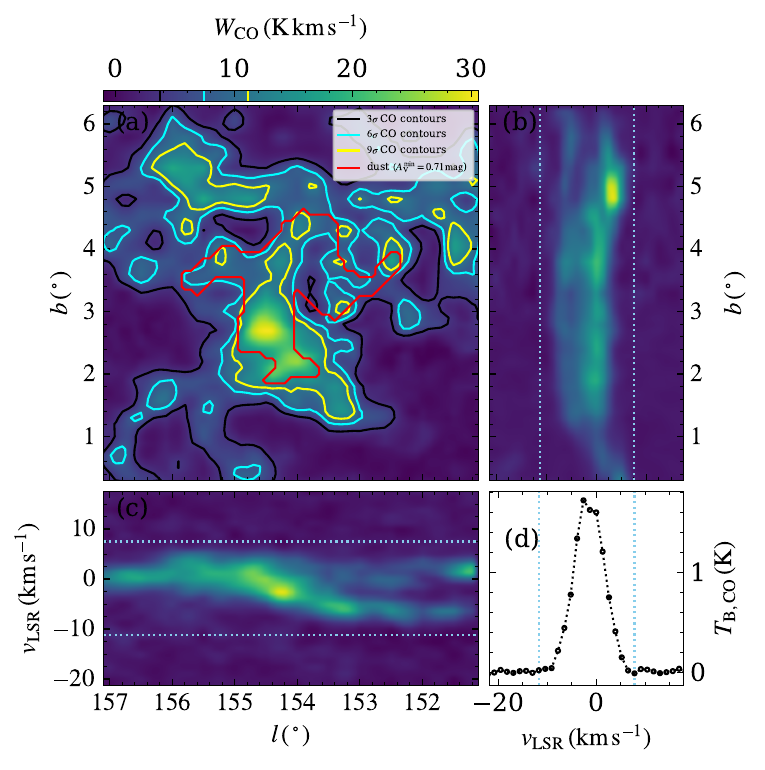}{0.33\textwidth}{(2) Cloud 205}
              \fig{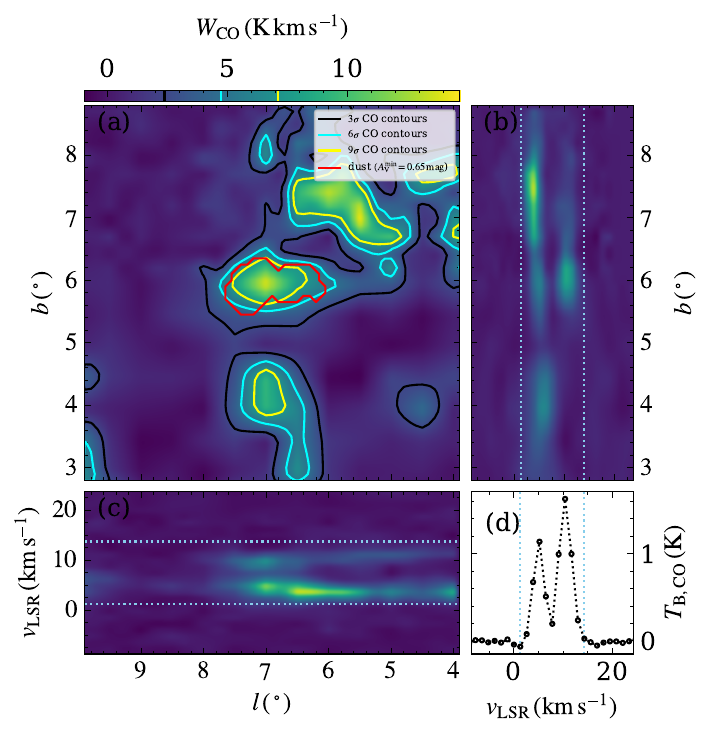}{0.33\textwidth}{(3) Cloud 221}}
    \caption{Morphological identification maps of clouds classified as possibly correlated dust-CO clouds, following a similar format to Figure~\ref{fig:dust-CO well correlated}.}
    \label{fig:dust-CO possibly correlated}
\end{figure*}

In contrast, clouds with multiple Gaussian components in their velocity-temperature profiles suggest line-of-sight overlaps with other clouds. These are categorized as ``possibly correlated dust-CO clouds.'' Figure~\ref{fig:dust-CO possibly correlated} provides examples of such MCs with potential correlations in dust and CO distributions. The left-hand and middle panels of Figure~\ref{fig:dust-CO possibly correlated}, depicting Cloud 158 and Cloud 205, show relatively weak CO emission within the dust contours. In the right-hand panel of Figure~\ref{fig:dust-CO possibly correlated}, illustrating Cloud 221, the superposition of two CO peaks along a single line-of-sight is evident. The velocity channel maps, displayed in Figure~\ref{fig:each_channel_MC_221}, indicate that the CO contours of both peaks are similar to the dust contour. Therefore, this cloud superimposed along a single line-of-sight cannot be resolved. As a result, these clouds are classified as possibly correlated dust-CO clouds.

Clouds whose CO emission line integrated intensity maps lack any discernible structure that aligns with dust contours are categorized as ``uncorrelated dust-CO clouds,'' exemplified by Figure~\ref{fig:each_channel_MC_31}.

For the strongly correlated dust-CO clouds, we fit the CO emission line profile with a single Gaussian function to determine the peak brightness temperature ($T_\mathrm{p}$), the corresponding velocity $v_\mathrm{LSR}$ ($v_\mathrm{p}$), and the velocity dispersion ($\sigma_v$).

The rms noise level, $\sigma_{W}$, in the velocity-integrated intensity map for each MC is estimated using the following equation:
\begin{equation}
\sigma_W = \frac{\sigma_\mathrm{c}}{\sqrt{N_\mathrm{channel}}},
\end{equation}
where $\sigma_\mathrm{c}$ (in units of $\mathrm{K \ km \ s^{-1}}$) represents the channel rms noise in the absence of emission, and $N_\mathrm{channel}$ denotes the number of independent channels within the total velocity width over which the integration is performed.

To quantify the dust mass in MCs, we use the average color excess $\overline{\Egk}$ from the MC catalog by \citet{2020MNRAS.493..351C}. This color excess is converted to visual extinction $A_{V}$ using the relation $\Egk/\Ebv = 2.15$ from \citet{2019MNRAS.483.4277C}, with an assumed total-to-selective extinction ratio $R_{V}$, yielding $A_{V} = R_{V} \cdot \overline{\Egk}/2.15$. For the diffuse ISM in the Milky Way, the typical $R_{V}$ is 3.1. However, \citet{2016ApJ...821...78S} revealed significant variations in $R_{V}$ across the Galactic plane through an analysis of the optical-infrared extinction curve. The distribution of $R_{V}$ from their study can be modeled by a Gaussian distribution with a mean of 3.32 and a standard deviation of 0.18.

As discussed in Section 5.2 of \citet{2023ApJS..269....6Z}, the variability of $R_{V}$ in the Galactic disk spans a broad range and appears on multiple scales, from small-scale structures within MCs to expansive kiloparsec regions. Notably, $R_{V}$ values within MCs range from approximately 2.6 to 3.3. On a broader scale, the $R_{V}$ distribution is well-characterized by a Gaussian function with a mean of 3.25 and a dispersion of 0.25. Further detailed discussions on $R_{V}$ can be found in Section~\ref{sec:challenges}. For this study, we adopt the results from the latter study and assign $R_{V}=3.25$ with a standard deviation of 0.25 to propagate uncertainties.

\section{Results and Discussion} \label{sec:results and discussions}

\begin{figure}[htbp]
    \centering
    \includegraphics[scale=.7]{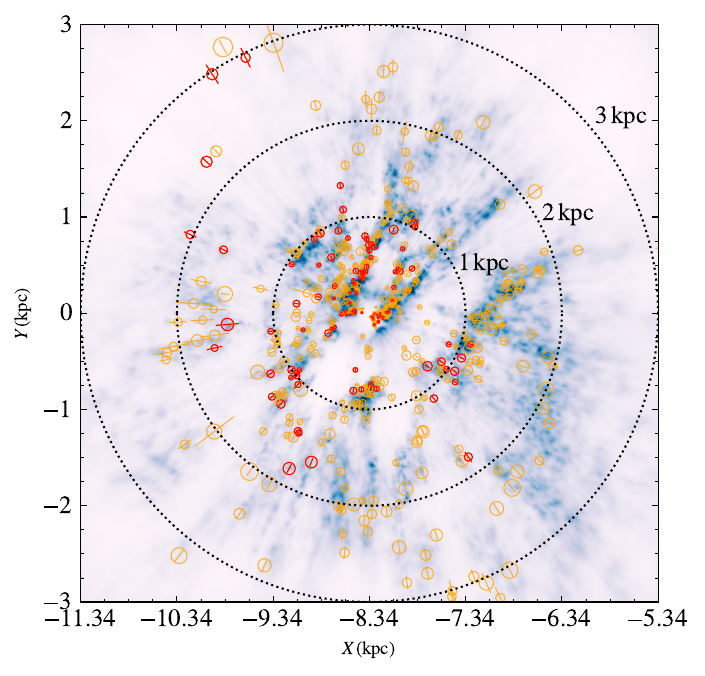}
    \caption{Spatial distribution of the classified MCs in the Galactic $X$-$Y$ plane, with the Galactic center at coordinates ($X$, $Y$) = (0, 0)\,kpc and the Sun at  ($X$, $Y$) =  ($-$8.34, 0)\,kpc. Concentric dotted circles denote distances of 1\,kpc, 2\,kpc, and 3\,kpc from the Sun. The 112 strongly correlated dust-CO clouds are denoted by red circles, while the others are represented by orange circles. The size of each circle is proportional to the cloud's physical size, and error bars indicate the uncertainty in distance. The underlying dust map is adapted from \citet{2019MNRAS.483.4277C}.}
    \label{fig:X-Y plane}
\end{figure}

Our investigation has successfully verified the existence of 112 strongly correlated dust-CO clouds, exhibiting consistent morphologies across both CO and dust observations. Additionally, 334 MCs have been tentatively identified as possibly correlated dust-CO clouds. However, for 24 clouds, no corresponding gas structures matching the dust distribution could be discerned. Moreover, due to the lack of CO data, 97 MCs could not be definitively classified. We have illustrated the spatial distribution of these different categories of MCs within the context of the Galactic large-scale structure in Figure~\ref{fig:X-Y plane}.

\subsection{Determining the Gas-to-Dust Ratio}

In this study, we have investigated the gas-to-dust ratio (GDR) in MCs as an application of our sample. The GDR quantifies the mass proportion of gas to dust in interstellar space and provides insight into the material conditions that promote star and planet formation. For our study, we have computed the average GDR value using the 112 strongly correlated dust-CO clouds. The total column density of hydrogen nuclei, denoted \NH, serves as a proxy for the gas mass, while the visual extinction, \Av, gauges the dust mass in MCs.

We estimate the column density of molecular hydrogen, \NHtwo, by leveraging a well-established correlation with the integrated intensity of the CO emission line, \Wco:
\begin{equation}
    \NHtwo = X_\mathrm{CO} \cdot \Wco \,,
\end{equation}
where $X_\mathrm{CO}$ represents the CO-to-$\Htwo$ conversion factor \citep[see][]{2013ARA&A..51..207B}. This factor, encoding the relative abundance of CO to \Htwo, is approximately $10^{-4}$ \citep{2015MNRAS.448.2187C, 2021ApJS..256....3P, 2013ARA&A..51..207B}. The conversion factor is typically valid for average properties over large scales, such as those of giant MCs ranging from 10 to 100\,pc \citep{1986ApJ...309..326D, 1991ARA&A..29..581Y, 1996ApJ...457..678B, 2000ApJ...541..142R,2002ApJ...579..270P}. The preferred $X_\mathrm{CO}$ value for the Milky Way disk is $2\times10^{20} \mathrm{\,cm^{-2}\,(K\, km\, s^{-1})^{-1}}$, with an uncertainty of about $\pm30\%$ \citep{2013ARA&A..51..207B}. For this analysis, we assume that $X_\mathrm{CO}$ is constant across all individual MCs and adopt the recommended value to compute $\NHtwo$.

To trace atomic gas, we utilize the $\HI$ 21\,cm emission line data from the HI4PI survey \citep{2016A&A...594A.116H}, which merges the Effelsberg-Bonn $\HI$ Survey \citep[EBHIS;][]{2016A&A...585A..41W} and the third Galactic All-Sky Survey \citep[GASS;][]{2015A&A...578A..78K}. Both EBHIS and GASS offer comparable angular resolution and sensitivity, culminating in the comprehensive HI4PI dataset that spans the entire sky with an angular resolution of $\theta_\mathrm{FWHM} = 16\arcmin.2$ (see Appendix \ref{sec:angular resolution} for a discussion on angular resolution) and LSR velocity bounds of $|v_\mathrm{LSR}|\leq 600\kms$ at a resolution of $\Delta v=1.29\kms$. The HI4PI data is publicly accessible. We apply the same resampling strategy to the HI4PI data that we used for the CO data, integrating the $\HI$ emission over the $v_\mathrm{LSR}$ range corresponding to the CO data to determine the $\HI$ column density via:
\begin{equation}
    \NHI=1.823\times10^{18}\int T_\mathrm{B,\scriptsize\HI}(v_\mathrm{LSR})\,\mathrm{d}v_\mathrm{LSR}\,,
\end{equation}
where $T_\mathrm{B,\scriptsize\scriptsize\HI}(v_\mathrm{LSR})$ represents the brightness temperature profile of the $\HI$ emission line, with $v_\mathrm{LSR}$ indicates the LSR velocity. It is crucial to note that this equation applies under the assumption of optically thin emission. In denser $\HI$ areas, especially at low Galactic latitudes where cold, low-temperature gas is common, $\HI$ 21\,cm line self-absorption can occur, meaning that the computed $\NHI$ represents a lower limit in such regions \citep{2016A&A...594A.116H}.

To infer the GDR using measured values of $\Wco$, $\NHI$, and $\overline{\Egk}$ from strongly correlated dust-CO clouds, we employ a hierarchical Bayesian approach. According to the Bayes' theorem, the posterior probability of the model parameters $\M$ given the data $\D$ is expressed as:
\begin{equation}\label{eq:Bayesian}
p(\M|\D)=\frac{p(\D|\M)p(\M)}{p(\D)} \propto p(\D|\M)p(\M|\Theta)p(\Theta),
\end{equation}
where $p(\D|\M)$ represents the \textit{likelihood} $\mathcal{L}(\M;\D)$, which is the probability of the observed data given the model parameters. The term $p(\M|\Theta)$ denotes the \textit{prior} distribution of the model parameters given the hyper-parameters, $p(\Theta)$ is the \textit{hyper-prior} distribution of the hyper-parameters, and $p(\D)$ is the prior predictive density, also known as the ``evidence.'' The evidence acts as a normalizing constant, allowing us to focus on the target distribution proportional to the posterior distribution $p(\M|\D)$. We estimate this posterior distribution using the No-U-Turn Sampler (NUTS), implemented in \texttt{PyMC} (v5.16.2\footnote{\url{https://doi.org/10.5281/zenodo.12724302}}) for Bayesian inference.

\begin{figure}[htbp]
\centering
\includegraphics[scale=.6]{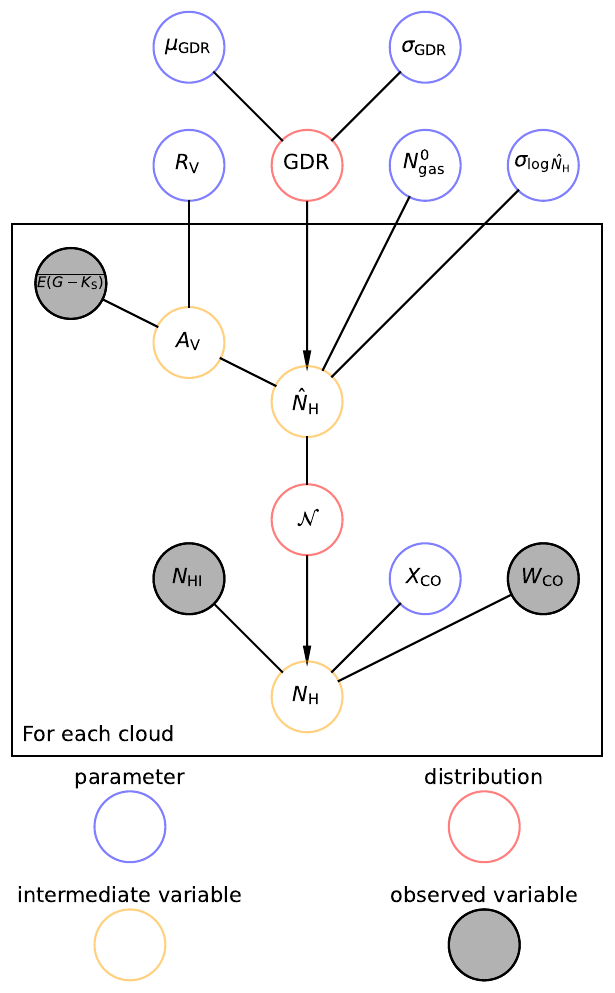}
\caption{Relationships among parameters in the hierarchical Bayesian model illustrated in this DAG.}
\label{fig:DAG}
\end{figure}

Figure~\ref{fig:DAG} illustrates our hierarchical Bayesian model using a directed acyclic graph (DAG). Below, we provide a detailed breakdown of each component.

For convenience, we define the hyper-parameters for which we conduct inference as $\Theta = \{\mu_\mathrm{GDR}, \sigma_\mathrm{GDR}, R_{V}, N_\mathrm{gas}^{0}, \sigma_{\log\hat{N}_\mathrm{H}}\}$. These hyper-parameters incorporate reasonable prior information about the model parameters. 
\begin{itemize}
    \item The mean GDR, $\mu_\mathrm{GDR}$, follows a uniform distribution: $\mu_\mathrm{GDR} \sim \mathrm{Uniform}(48, 51)$.
    \item The standard deviation of GDR, $\sigma_\mathrm{GDR}$, follows a Half-Cauchy distribution: $\sigma_\mathrm{GDR} \sim \mathrm{Half-Cauchy}(1)$.
    \item Given $\mu_\mathrm{GDR}$ and $\sigma_\mathrm{GDR}$, the GDR itself is drawn from a log-normal distribution: $p(\GDR|\mu_\mathrm{GDR}, \sigma_\mathrm{GDR}) \sim \mathrm{Log-Normal}(\mu_\mathrm{GDR}, \sigma_\mathrm{GDR})$.
    \item The total-to-selective extinction ratio $R_{V}$, as described in Section~\ref{sec:method}, follows a normal distribution: $R_{V} \sim \mathrm{Normal}(3.25, 0.25)$.
    \item The background gas, $N_\mathrm{gas}^{0}$, follows a log-normal distribution: $N_\mathrm{gas}^{0} \sim \mathrm{Log-Normal}(47.5, 1)$.
    \item The standard deviation of the expected hydrogen column density in the logarithmic space, $\sigma_{\log\hat{N}_\mathrm{H}}$, follows a Half-Cauchy distribution: $\sigma_{\log\hat{N}_\mathrm{H}} \sim \mathrm{Half-Cauchy}(1)$.
\end{itemize}
The overall hyper-prior distribution is the product of the individual distributions: $p(\Theta) = p(\mu_\mathrm{GDR}) p(\sigma_\mathrm{GDR}) p(R_{V}) p(N_\mathrm{gas}^{0}) p(\sigma_{\log\hat{N}_\mathrm{H}})$.

The model parameters are expressed as $\M = \{\hat{N}_\mathrm{H} = \GDR \cdot \Av + N_\mathrm{gas}^{0}\}$. The visual extinction $\Av$ is derived using the extinction coefficient $R_{V}$ and the measured color excess $\overline{\Egk}$, yielding $\Av = R_{V} \cdot \overline{\Egk} / 2.15$. Given the hyper-parameters $\Theta$, the prior distribution of the model parameters follows a log-normal distribution: $p(\M|\Theta) \sim \mathrm{Log-Normal}(\log{\hat{N}_\mathrm{H}}, \sigma_{\log\hat{N}_\mathrm{H}})$.

The measured parameters are represented as $\D = \{N_\mathrm{H} = \NHI + 2 \Xco \cdot \Wco\}$. Given the model parameters $\mathcal{M}$, the likelihood function follows a normal distribution: $p(\D|\M) \sim \mathrm{Normal}(\M, \sigma_\mathcal{D})$, where the observed measurement error $\sigma_\mathcal{D}$ is propagated by the quadratic sum of errors: $\sigma_\mathcal{D}^{2} = \sigma_{N_{\scriptsize\HI}}^{2} + (2X_\mathrm{CO}\sigma_{W_\mathrm{CO}})^2$.

\begin{figure}[htbp]
\centering
\includegraphics[scale=.6]{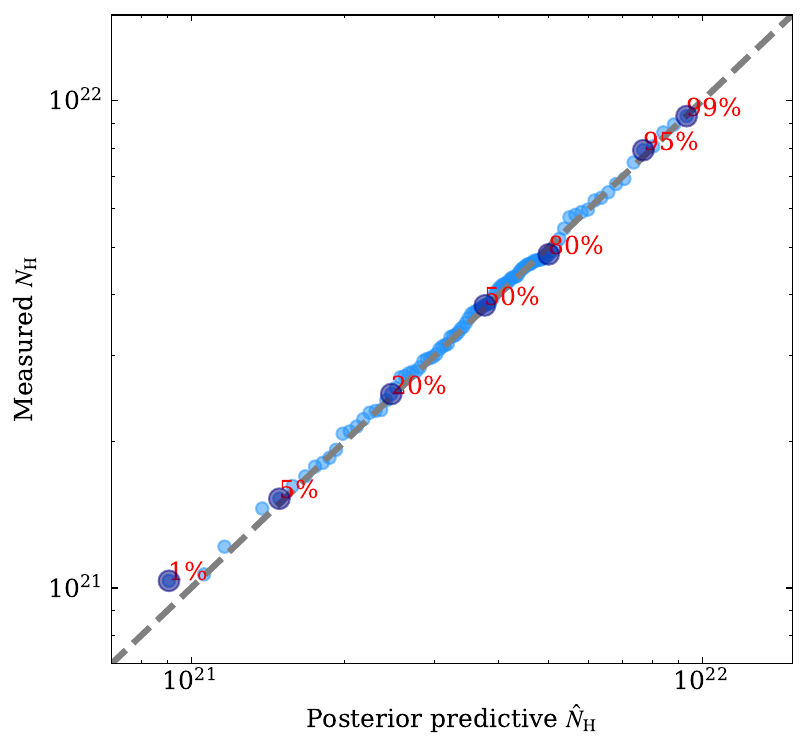}
\caption{Quantile-Quantile ($Q$-$Q$) plot comparing posterior predictive $\hat{N}_\mathrm{H}$ and measured $N_\mathrm{H}$. If the predictive and measured data followed the same distribution, the quantiles would align along the 1:1 line. Significant discrepancies are observed below the $\sim$ 2nd percentile.}
\label{fig:Q-Q}
\end{figure}

These components collectively form the Bayesian inference framework for our model, enabling statistical analysis and parameter estimation based on observed data. After deriving the posterior distribution, we evaluate the adequacy of our model following the methodology outlined in \citet{2023arXiv230204703E}. By conducting posterior predictive checks, we compare the quantiles from the posterior predictive $\hat{N}_\mathrm{H}$ with those from the measured $N_\mathrm{H}$, as shown in Figure~\ref{fig:Q-Q}. Although discrepancies exist, especially in the lower tail of the distribution, our model generally captures most properties of the measured data.

\subsection{Best-fit GDR Estimated from the 112 Strongly Correlated Dust-CO Clouds}

\begin{figure}[htbp]
    \centering
    \includegraphics[scale=1.]{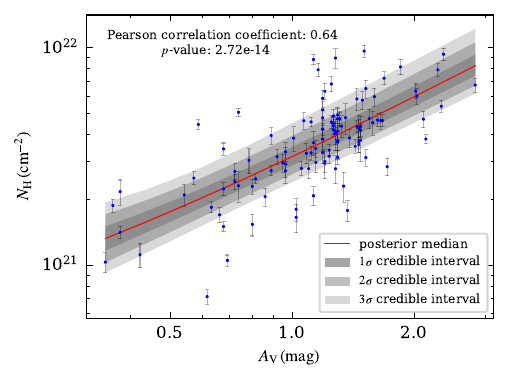}
    \caption{Best-fit correlation between the column density of hydrogen nuclei $N_\mathrm{H}$ and visual dust extinction $\Av$ for the 112 strongly correlated dust-CO clouds, yielding an average GDR value of $(2.80_{-0.34}^{+0.37})\times10^{21}\,\mathrm{cm^{-2}\,mag^{-1}}$. The best-fit GDR is represented by the red line, while the corresponding different credible intervals are depicted by the gray shading.}
    \label{fig:Best_fitting_GDR}
\end{figure}

The relationship between the column density of hydrogen nuclei ($N_\mathrm{H}$) and visual dust extinction ($A_{V}$) for the 112 strongly correlated dust-CO clouds we classified is depicted in Figure~\ref{fig:Best_fitting_GDR}. We have superimposed the best-fit line, representing the GDR, onto the same figure. The ensemble-averaged GDR for these clouds is $\GDR = (2.80_{-0.34}^{+0.37}) \times 10^{21} \, \mathrm{cm^{-2} \, mag^{-1}}$, with a background gas column density of $N_\mathrm{gas}^0 = (3.32^{+2.13}_{-1.63}) \times 10^{20} \, \mathrm{cm^{-2}}$. The fit aligns well with the observed data, indicating that it robustly represents the underlying correlation. A Pearson correlation test yielded a coefficient of 0.64 with a $p$-value of $2.72 \times 10^{-14}$, signifying a significant and moderately strong correlation.

The relationships among velocity-integrated CO intensity ($W_\mathrm{CO}$), average color excess ($\overline{\Egk}$), and $\HI$ column density ($\NHI$) are shown in the first row of Figure~\ref{fig:properties}. The results of the Pearson correlation tests are indicated on the plots. It is evident that there is a moderate correlation between $W_\mathrm{CO}$ and $\overline{\Egk}$, while the correlations between $W_\mathrm{CO}$ and $\NHI$, as well as between $\overline{\Egk}$ and $\NHI$, are comparatively weaker.

One cloud, identified as Cloud 36, exhibits a relatively low $W_\mathrm{CO} \sim 0.47 \, \mathrm{K \, km \, s^{-1}}$, significantly deviating from the sample values, as shown in panels (a) and (b) of Figure~\ref{fig:properties}. However, its high $\NHI \sim 5.27 \times 10^{20} \, \mathrm{cm^{-2}}$ and $\overline{\Egk} \sim 0.41 \, \mathrm{mag}$ suggest that Cloud 36 may possess unique physical conditions or be in a different evolutionary stage. Upon re-examining the identification process for this cloud, we confirmed that it meets our criteria for strongly correlated dust-CO clouds. Therefore, it is retained in our sample.

For context, \citet{1977ApJ...216..291S} and \citet{1978ApJ...224..132B} obtained a GDR of $1.87 \times 10^{21}\pscmpmag$ from UV absorption and stellar reddening measurements toward early-type stars, a value indicative of the diffuse ISM within the Galactic plane. \citet{1995A&A...293..889P} conducted an analysis of the diffuse X-ray halos surrounding 25 point sources and 4 supernova remnants (SNRs) through soft X-ray scattering, resulting in a derived ratio of $\GDR=1.79\times10^{21}\pscmpmag$. Our relatively higher GDR value reflects the denser gas and dust environments in MCs within the Galactic plane. The GDR values extracted from the recent literature are listed in Table~\ref{tab:gas-to-dust ratio} for comparison. Although different studies used various dust and/or gas tracers and investigated different regions, our findings are generally consistent with the values listed in the table.
\input{gas-to-dust_ratio}

\subsection{Variability of GDR across Individual MCs}

\begin{figure}[htbp]
    \centering
    \includegraphics[scale=1.]{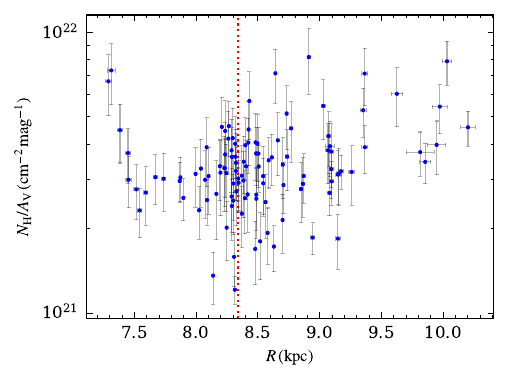}
    \caption{Variation of $N_\mathrm{H}/A_{V}$ for the individual clouds located at different Galactocentric distances $R$. The Sun's location at $R\sim8.34\,\mathrm{kpc}$ is indicated by the red dotted line for reference.}
    \label{fig:GDR-R}
\end{figure}

Despite the clear linear trends observed, we note a non-negligible scatter in the $N_\mathrm{H} - A_{V}$ relationship for individual MCs. This dispersion underscores the heterogeneity in the GDR values, which are further explored as a function of Galactocentric distance, $R$, in Figure~\ref{fig:GDR-R}. GDRs for individual MCs are found to be elevated beyond the ensemble-averaged GDR at distances $R \gtrsim 8.5\,\mathrm{kpc}$, particularly in regions such as the Galactic anti-center and the Perseus Arms. This is likely attributable to the abundance of cold gas in the outer Galactic disk, where molecular cloud density and star formation rates are lower, leading to less metal enrichment \citep{2012A&A...543A.103P, 2021ApJ...910...46L}. Furthermore, MCs located within the Sagittarius Arm ($R \lesssim 7.5\,\mathrm{kpc}$) display GDRs that exceed those estimated in the solar neighborhood ($R\sim8.34\,\mathrm{kpc}$). This physical picture is also reflected in the variations of $\Wco$, $\NHI$, and $\overline{\Egk}$ with Galactocentric distance $R$, respectively, as shown in the second row of Figure~\ref{fig:properties}. The variations of $N_\mathrm{H}/\Av$ with Galactic longitude and latitude are also shown in the last two rows of Figure~\ref{fig:properties}, respectively.

\begin{figure}[htbp]
    \centering
    \includegraphics[scale=.6]{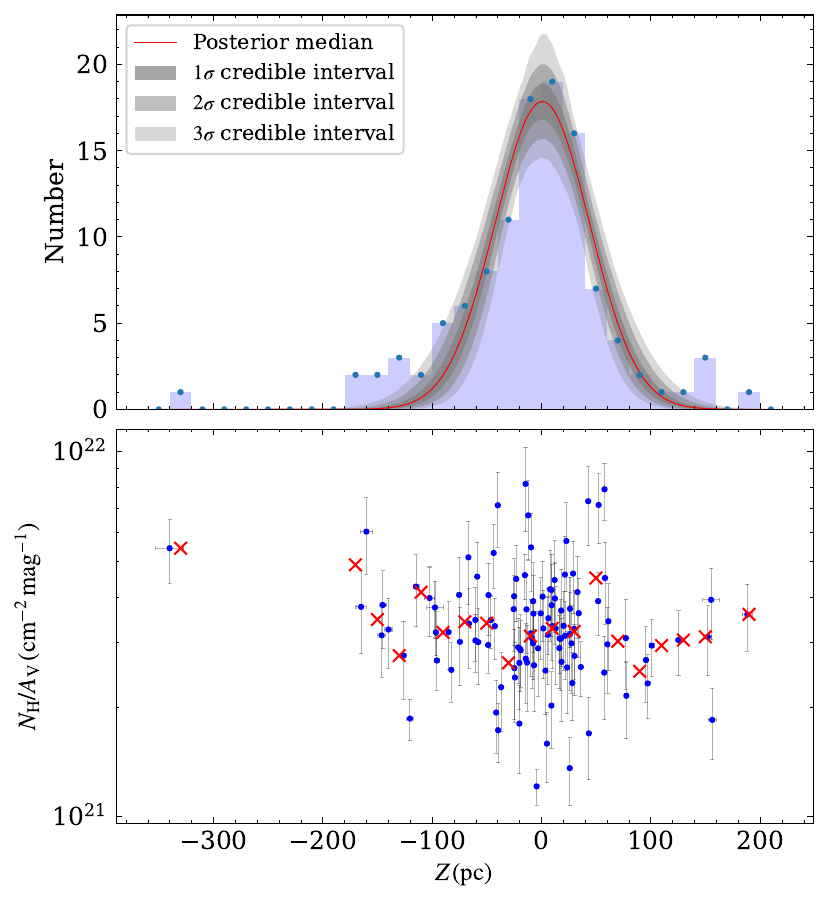}
    \caption{\textit{Upper panel}: Histogram of the vertical distances from the Galactic plane $Z$, where the red line represents the posterior median, and the shaded areas represent the $1\sigma$, $2\sigma$, and $3\sigma$ credible intervals for the vertical distribution fit of the MCs, respectively. \textit{Lower panel}: The dependence of $N_\mathrm{H}/A_{V}$ on the individual clouds at different $Z$. The median $N_\mathrm{H}/A_{V}$ values for discrete $Z$ bins are marked with red crosses, with each bin spanning 20 pc.}
    \label{fig:GDR-Z}
\end{figure}

Yet, as depicted in the lower panel of Figure~\ref{fig:GDR-Z}, the GDR does not exhibit significant fluctuations with respect to the perpendicular distance from the Galactic plane, $Z$, when the data is organized into 20 pc bins.

\subsection{Scale Height of MCs}
Based on the distribution of $Z$, we investigated the scale height of MCs. We assume that the vertical distribution of clouds follows a Gaussian profile, given by
\begin{equation}
n(Z) \propto \exp{\left[-\frac{(Z-Z_{0})^2}{2h_{Z}^2}\right]},
\end{equation}
where $h_{Z}$ is the scale height of MCs, and $Z_{0}$ is the offset from the Galactic symmetry mid-plane. The histogram of the vertical distances of MCs from the Galactic plane ($Z$) is shown in the upper panel of Figure~\ref{fig:GDR-Z}. The best-fit results yield a scale height of $h_{Z} = 43.3_{-3.5}^{+4.0}$ pc and an offset of $Z_{0} = 0.5 \pm 2.9$ pc.

MCs are the primary sites for star formation, and their connection to stellar clusters sheds light on key processes in star formation and evolution within the Galaxy. Stellar clusters form within MCs, and over time these young stars gradually move away from their parent clouds. Understanding the age and position of these clusters is crucial for studying the structure and evolution of the Milky Way. \citet{2020A&A...640A...1C} estimated the age, distance modulus, and interstellar extinction for approximately 2000 clusters based on \textit{Gaia} photometry and their mean \textit{Gaia} parallax. This methodology yielded a sample of 1867 clusters with reliable parameters. When projected onto the Galactic plane, the positions of young clusters generally align with the expected spiral pattern, particularly along the Sagittarius, Local, and Perseus Arms \citep[see][]{2019AJ....158..122K,2020AJ....160..279K,2020A&A...640A...1C,2021A&A...652A.162C,2022A&A...661A.118C}. Notably, the majority of MCs in our sample are located within these spiral arms \citep[see Figure~7 of][]{2020MNRAS.493..351C}.

Numerous studies have documented the well-known dependency of cluster scale height on age and/or Galactocentric distance \citep[e.g.,][]{2016AA...593A.116J,2019AJ....158..122K,2020AJ....160..279K}. Younger open clusters (OCs) (age $< 10$\,Myr) distinctly outline the spiral arms in the solar vicinity \citep{2023AJ....166..170J}. As these clusters age, they begin to occupy regions between the spiral arms. Typically, OCs depart from their birth sites after $\sim10-20$\,Myr, and most OCs aged $\sim20-30$\,Myr subsequently populate the inter-arm regions, where they remain for the duration of their existence. Therefore, we expect that the vertical distribution of MCs may be consistent with that of young OCs.

Table~\ref{tab:scale_height} lists the scale heights of different Galactic disk components for comparison. Although the Galactic disk is a complex, multi-component system consisting of gravitationally coupled stars and the ISM within the potential of a dark matter halo \citep{2022A&A...665A..23S}, and models vary (such as Gaussian disk, exponential disk, and self-gravitating isothermal disk), our result for the scale height remains consistent with that of the young stellar population within the margin of error. Additionally, our findings align with theoretical simulations \citep[$h_{Z}\sim40-60$ pc at the Galactocentric distance of the Sun;][]{2022MNRAS.515.1663J,2024arXiv240614661M}. Overall, these results support our approach to effectively identify correlated dust-CO clouds.
\input{scale_height.tex}

\subsection{Challenges in Estimating the GDR in MCs}\label{sec:challenges}

The column density of molecular gas in MCs is often estimated using emission lines from low-$J$ rotational transitions of CO, which are then scaled to the assumed column density of molecular hydrogen ($\NHtwo$) \citep{2010ApJ...716.1191W}. This method assumes that CO completely traces the spatial distribution of $\Htwo$ \citep{2022ApJ...933..179O}. However, a fraction of $\Htwo$ in MCs cannot be traced by CO due to CO's lower capability to shield itself from far-UV radiation compared to $\Htwo$. Consequently, the region where $\mathrm{C}^+$ transforms into CO is closer to the cloud core than the region where $\HI$ transforms into $\Htwo$ \citep[See also Figure~31.2 of][]{2011piim.book.....D}. \citet{2012A&A...543A.103P} identified the transitions of $\HI$ to $\Htwo$ and $\Htwo$ to CO at visual extinctions of $A_{V} \approx 0.2$ mag and $A_{V} \approx 1.5$ mag, respectively, consistent with theoretical models predicting dark-$\Htwo$ gas. \citet{2020A&A...639A..26K} found that the transition from $\HI$ to $\Htwo$ does not significantly affect dust properties in the diffuse ISM. Although $\HI$ emission decreases with $\Htwo$ formation, the associated optical extinction remains unchanged. Dust may be heated by stellar radiation within the Galactic disk \citep{2001ApJ...547..792D}. Dust temperature, crucial for molecular gas formation, inversely correlates with total gas column density. Heating and cooling processes in the ISM vary significantly with changes in volume density \citep[e.g.,][]{2003ApJ...587..278W}. Warm dust is indicative of the outer, primarily atomic regions of clouds, while cold dust is associated with the inner, primarily molecular regions. This variability in dust temperature may also contribute to the scatter observed in GDR \citep{2012A&A...548A..22M,2024A&A...682A.161S}.

\citet{2010ApJ...716.1191W} developed a model for PDRs within individual spherical clouds, estimating that under standard Galactic conditions, dark gas (DG) constitutes approximately 30\% ($f_\mathrm{DG} \sim 0.3$) of the total molecular gas mass. This estimate was considered relatively unaffected by variations in cloud and environmental properties. However, observational studies suggest that $f_\mathrm{DG}$ can vary under different environmental conditions \citep{2012A&A...543A.103P,2022ApJ...933..179O}. The DG component has been indirectly detected using other tracers such as gamma rays from cosmic-ray interactions with gas, FIR/submillimeter dust continuum emission \citep{2010ApJ...716.1191W}, and ionized and neutral carbon \citep{2020A&A...643A.141M}. Within the Galactic plane ($|b| \leq 10^\circ$), isolating signals from individual clouds is challenging due to the potential blending of emissions from multiple overlapping MCs \citep[e.g.,][]{2020A&A...639A..26K}.

\citet{2017MNRAS.471L..52T} conducted numerical simulations of dusty, supersonic turbulence within MCs, revealing that both small ($0.1\,\mu\mathrm{m}$) and large ($\gtrsim1\,\mu\mathrm{m}$) dust grains trace the large-scale structure of the gas. These simulations indicate that turbulence causes the agglomeration of larger dust grains in denser cloud areas, potentially leading to ``coreshine'' phenomena in dark clouds.

The total-to-selective extinction ratio $R_{V}$, which describes extinction curves, also provides crucial insights into dust grain properties. The relationship between larger dust grain sizes and higher $R_{V}$ values was demonstrated using the model developed by \citet{2001ApJ...548..296W}. It is commonly understood that sight lines traversing dense MCs with high extinction are likely to display higher $R_{V}$ values, often explained by dust grain growth through accretion and coagulation processes \citep{2024A&A...682A.161S}. Given the substantial mass contribution from very large dust grains, which are not significantly detected by optical extinction \citep{1979ARA&A..17...73S}, the GDR derived from extinction measurements could be overestimated.

A dust grain in a dense MC encounters a markedly distinct radiation field and collision rate compared to one in a diffuse atomic cloud \citep{2016ApJ...821...78S}. When subjected to radiation, dust grains are increasingly prone to destruction through mechanisms such as sputtering from incoming atoms or ions, photolysis induced by UV photons, and photodesorption on their surfaces \citep{2003ARA&A..41..241D}. Both growth and reduction mechanisms in dust grains tend to balance each other out, leading to a decrease in average grain size within extinction regions of about $0.3 < \Ebv < 1.2$ \citep{2023ApJS..269....6Z}. This behavior in grain size distribution might be attributed to the predominance of reduction mechanisms in these regions, resulting in lower $R_{V}$ values within MCs. Variations in $R_{V}$ could also be influenced by the chemical composition of dust grains and dust temperature. \citet{2023ApJS..269....6Z} conducted a detailed investigation into the relationships between $R_{V}$ and various parameters, such as dust temperature, the spectral index of dust emissivity, column densities, ratios of atomic to molecular hydrogen, and GDR, observing variability depending on different levels of extinction, which reflects the variation in the properties and evolution of dust grains.

The CO-to-$\Htwo$ conversion factor ($X_\mathrm{CO}$) varies depending on the properties of the ISM, including metallicity, density, temperature, and the intensity of the UV radiation field, as demonstrated by theoretical simulations  \citep{2017ApJ...843...38G,2018ApJ...858...16G}. \citet{2022ApJ...931....9L} reported that the $X_\mathrm{CO}$ factor exhibits significant variability on the parsec scale, indicated by a broad log-normal-like frequency distribution. This variability suggests that CO may not reliably trace the $\Htwo$ column density at sub-cloud scales. Nonetheless, observations of local ISM clouds indicate that the $X_\mathrm{CO}$ factor remains relatively constant, with deviations usually within a factor of 2 \citep{2013ARA&A..51..207B}. In this paper, we estimated $\NHtwo$ using the recommended values provided by \citet{2013ARA&A..51..207B}. For a detailed discussion of $X_\mathrm{CO}$, we refer the reader to \citet{2013ARA&A..51..207B}.

The examination of dust and gas in non-star-forming high-latitude MCs is beneficial due to the less complex physical environment, as noted by \citet{2022A&A...668L...9M}. In contrast, the Galactic plane is replete with star-forming regions, where star formation and stellar feedback significantly alter the physical conditions, complicating the analysis of gas and dust properties \citep{2024MNRAS.529.1091Z}. Additionally, the intricate structure of clouds and projection effects can further complicate the measurement of physical parameters in MCs \citep{2018MNRAS.474.4672L, 2024ApJ...961..153C}.

$\HI$ narrow-line self-absorption (HINSA) features are prominent indicators of cold $\HI$ intermixed with $\Htwo$, because they align with CO emissions and present line widths similar to or narrower than those of CO \citep{2003ApJ...585..823L}. MCs numbered 71, 90, 95, 139, 143, 191, 210, 234, 267, 311, 370, 450, 509, and 553 exhibit notable HINSA characteristics, underlining the utility of HINSA in identifying regions where cold $\HI$ coexists with $\Htwo$.

To sum up, further investigation is required to improve our understanding of the interactions and dynamics between gas and dust tracers in the MCs within the Galactic plane.

\section{Summary} \label{sec:summary}

Utilizing 3D gas \citep{2001ApJ...547..792D, 2016A&A...594A.116H} and dust extinction data \citep{2019MNRAS.483.4277C, 2020MNRAS.493..351C}, we have identified 112 strongly correlated dust-CO clouds, which are well associated with both dust and CO, in regions of low Galactic latitude ($|b|\leq10^\circ$). For the subset of 112 strongly correlated dust-CO clouds, we have quantified their physical properties using data derived from CO observations and generated a catalog. The catalog includes each MC's Galactic coordinates \((l, b)\), distance \(d_{0}\), boundary extinction $A_{V}^\mathrm{min}$, visual extinction \(A_{V}\), peak CO emission line brightness temperature \(T_\mathrm{p}\), peak CO emission line brightness LSR velocity \(v_\mathrm{p}\), LSR velocity dispersion \(\sigma_v\), LSR velocity range for the integrated intensity \(\Delta V\), and total intensity of the CO emission line \(W_\mathrm{CO}\). We have also detailed the measured $\HI$ column densities $\NHI$, the column density of hydrogen nuclei $N_\mathrm{H}$, and the ratio $N_\mathrm{H}/A_{V}$ in our public database. The database containing these parameters and CO identification maps (such as Figure~\ref{fig:dust-CO well correlated}) for these MCs is publicly accessible through the website doi:\dataset[10.12149/101367]{\doi{10.12149/101367}}.

We have observed a linear correlation between the gas and dust contents within these MCs, albeit with notable scatter. We postulate that this dispersion is likely due to the diverse and complex physical processes occurring within the individual MCs, potentially impacting the GDR. The average GDR across our sample is found to be $\GDR = (2.80_{-0.34}^{+0.37})\times10^{21}\,\mathrm{cm^{-2}\,mag^{-1}}$, aligning with values presented in prior studies. While the GDR of individual MCs tends to be higher both inside $R\lesssim7.5\,\mathrm{kpc}$ and beyond $R\gtrsim8.5\,\mathrm{kpc}$ compared to the overall average GDR of clouds, we found no significant variation in the median GDR values of MCs when categorized by their distances from the Galactic plane.

We have derived the scale height of strongly correlated dust-CO clouds, $h_{Z} = 43.3_{-3.5}^{+4.0}$\,pc, which is in excellent agreement with that of the young stellar population. This finding validates our methodology for effectively identifying correlated dust-CO clouds.

\begin{acknowledgments}
The authors would like to express their gratitude to the anonymous referee for their valuable comments and suggestions, which contributed to the improvement of the manuscript. This work is partially supported by the National Key R\&D Program of China No. 2019YFA0405500, the National Natural Science Foundation of China 12173034 and 12322304, the Innovation and Entrepreneurship Training Projects for College Student of Yunnan University 202010673028, and Yunnan University grant No. C619300A034. We acknowledge the science research grants from the China Manned Space Project with NO. CMS-CSST-2021-A09, CMS-CSST-2021-A08, and CMS-CSST-2021-B03.
\end{acknowledgments}

\vspace{5mm}\,
\facilities{Effelsberg, Parkes}


\software{Astropy \citep{2013A&A...558A..33A, 2018AJ....156..123A, 2022ApJ...935..167A},
          Matplotlib \citep{Hunter:2007},
          PyMC \citep{pymc2023},
          OpenCV-Python \citep{opencv-python},
          SciPy \citep{2020SciPy-NMeth},
          NumPy \citep{harris2020array},
          astrodendro \citep{2008ApJ...679.1338R},
          reproject (\url{https://reproject.readthedocs.io/}),
          daft (\url{https://docs.daft-pgm.org/en/latest/})
}




\appendix
\restartappendixnumbering

\section{Angular Resolution of Gas Data}\label{sec:angular resolution}

\begin{figure}[htbp]
\centering
\includegraphics[scale=.54]{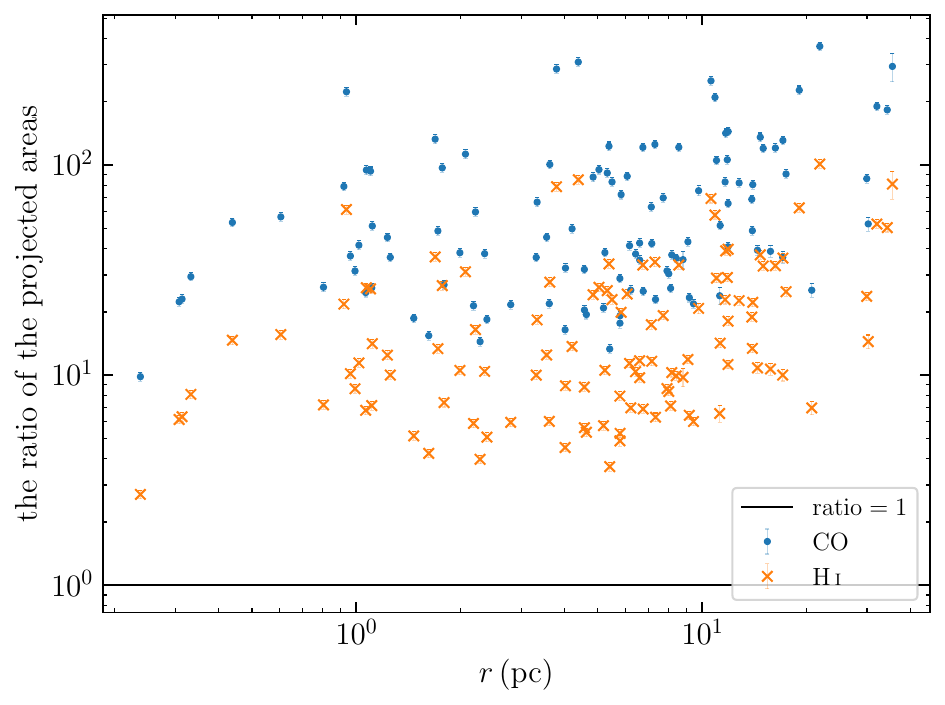}
\caption{Ratio of the projected areas versus radius. For each MC, blue dots represent the area ratio estimated using the CO data angular resolution, while orange crosses represent the ratio using the $\HI$ data angular resolution. Points above the horizontal line (ratio = 1) indicate that the projected area of the MC is larger than the single pixel area. The uncertainties in the area ratio are due to distance uncertainties.}
\label{fig:area-ratio}
\end{figure}

The angular resolution of the CO data used in this study is approximately $8\arcmin.5$, whereas the $\HI$ data has an angular resolution of $16\arcmin.2$. To determine the suitability of these resolutions for our analysis, we combined the physical radius ($r$) and distance ($d_{0}$) information of MCs from the catalog by \citet{2020MNRAS.493..351C}. We calculated the projected area of each MC ($\pi r^{2}$) and compared it to the area of a pixel at the given angular resolution and distance of the MC. This ratio of projected areas was then analyzed to assess the impact of angular resolution on our measurements. As shown in Figure~\ref{fig:area-ratio}, the area ratio exceeds 1 for all MCs strongly correlated with dust-CO, indicating that the angular resolution of the gas data does not significantly affect our gas content measurements.

\section{Append Figures}
\begin{figure*}[htbp]
\centering
\includegraphics[scale=.6]{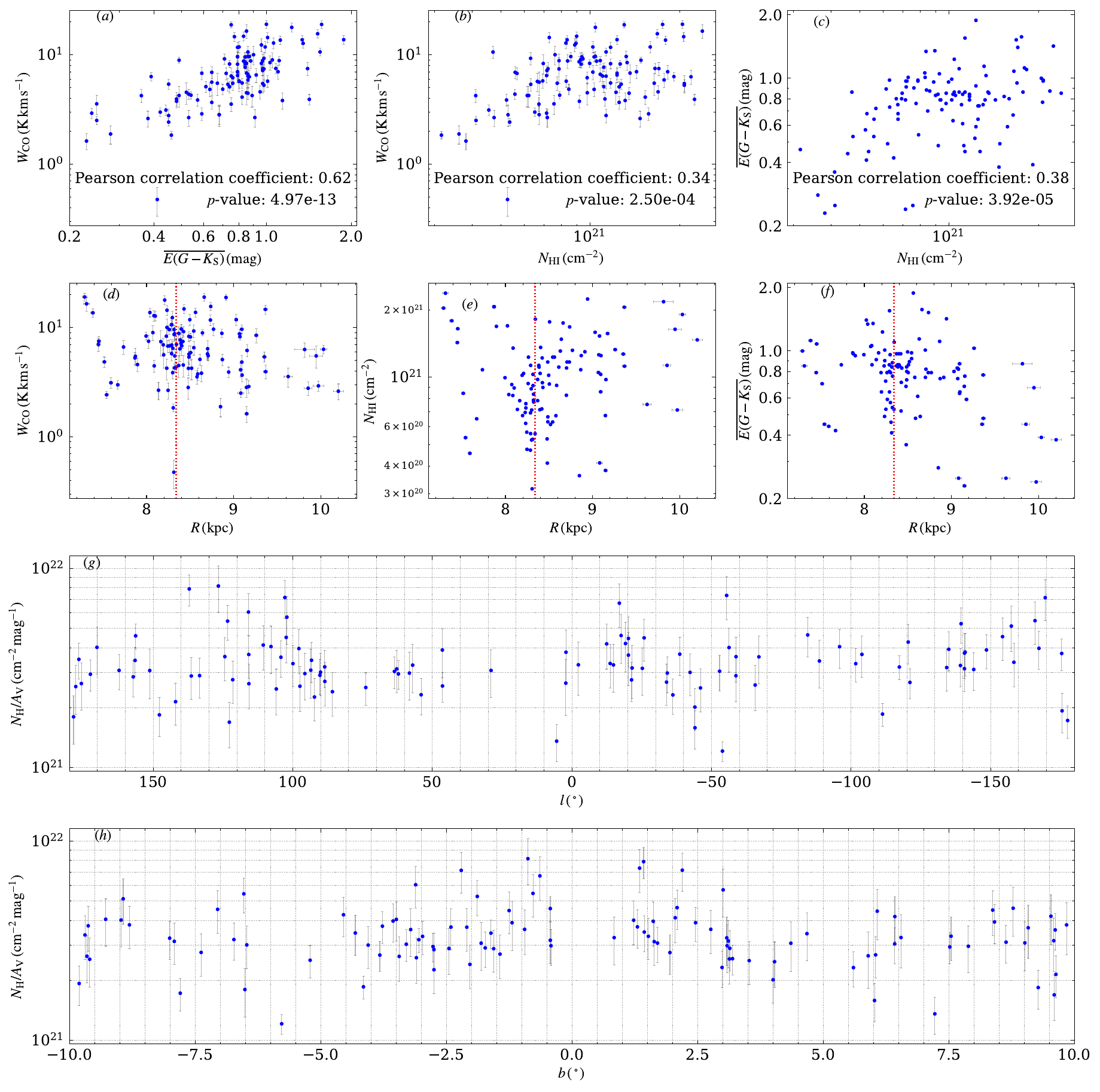}
\caption{Relations between various physical properties of MCs. Panel ($a$) shows the velocity-integrated CO intensity ($\Wco$) vs. average color excess ($\overline{\Egk}$) with a Pearson correlation coefficient of 0.62 and a $p$-value of $4.97\times10^{-13}$. Panel ($b$) displays $\Wco$ vs. $\HI$ column density ($\NHI$) with a Pearson correlation coefficient of 0.34 and a $p$-value of $2.50\times10^{-4}$. Panel ($c$) illustrates $\overline{\Egk}$ vs. $\NHI$ with a Pearson correlation coefficient of 0.38 and a $p$-value of $3.92\times10^{-5}$. Panel ($d$) presents $\Wco$ vs. Galactocentric distances ($R$). The Sun's position at $R \sim 8.34 \mathrm{\,kpc}$ is marked by the red dotted line for reference. The subsequent panels ($e$) and ($f$) use the same position reference of the Sun. Panel ($e$) shows $\NHI$ vs. $R$. Panel ($f$) depicts $\overline{\Egk}$ vs. $R$. Panel ($g$) examines the ratio $N_\mathrm{H} / \Av$ vs. Galactic longitude ($l$). Panel ($h$) investigates the ratio $N_\mathrm{H} / \Av$ vs. Galactic latitude ($b$).}
\label{fig:properties}
\end{figure*}

The relationships between various physical properties of MCs that exhibit strong dust-CO correlations are presented in Figure~\ref{fig:properties}.







\bibliographystyle{aasjournal}
\bibliography{main}{}



\listofchanges


\end{CJK*}
\end{document}

%% file: gas-to-dust_ratio.tex
\startlongtable 
\begin{deluxetable*}{lcl}
    \tablecolumns{3}
    \tabletypesize{\scriptsize}
    \setlength{\tabcolsep}{0.05in}
    \tablewidth{0pt}
    \tablecaption{Comparisons of $N_\mathrm{H}/\Av$ from the Literature Values}
    \tablehead{
    \colhead{Reference} &
    \colhead{$N_\mathrm{H}/\Av$} &
    \colhead{Comment} \\
    \colhead{} &
    \colhead{($10^{21}\pscmpmag$)} &
    \colhead{} 
    }
    \startdata
This study & $2.80_{-0.34}^{+0.37}$ & $\HI$ and CO emission lines from 112 strongly correlated dust-CO clouds within $|b|\leq10^\circ$\\
\citet{2009MNRAS.400.2050G} & $2.21\pm0.09$ & X-ray absorbing column densities toward 22 SNRs with $\Ebv\lesssim10\mathrm{\,mag}$\\
\citet{2016ApJ...826...66F} & $2.87\pm0.12$ & X-ray absorbing column densities toward 17 SNRs with $\Ebv\lesssim10\mathrm{\,mag}$\\
\citet{2014ApJ...780...10L} & $2.68$ & Only $\HI$ emission line from Galactic latitudes $|b|\gtrsim20^\circ$ and $\Ebv\lesssim0.1\mathrm{\,mag}$\\
\citet{2015MNRAS.448.2187C} & $2.41 \pm 0.01$ & $\HI$ and CO emission lines from the Galactic anti-center $|b|\gtrsim10^\circ$ and $\Ebv\lesssim1\mathrm{\,mag}$ \\
\citet{2017MNRAS.471.3494Z} & $2.47\pm0.04$ & X-ray absorbing column densities toward 19 SNRs and 29 X-ray binaries \\
\citet{2017ApJ...846...38L} & $2.84$ & Only $\HI$ emission line in the low-column-density regime with $\Ebv\approx45\mathrm{\,mmag}$ \\
\citet{2018ApJ...862...49N} & $3.03\pm0.52$ & $\HI$ absorption line from purely atomic sight lines at $|b|>5^\circ$ \\
\citet{2024ApJ...961..204S} & $3.32 \pm 0.13$ & $\HI$ emission, $\Htwo$ far-UV absorption lines toward 51 quasars and gas at $|v_\mathrm{LSR}|\leq 600\kms$ \\
\citet{2024ApJ...961..204S}& $2.97 \pm 0.10$ & $\HI$ emission, $\Htwo$ far-UV absorption lines toward 51 quasars and gas at $|v_\mathrm{LSR}|\leq 90\kms$ \\
\enddata
\tablecomments{For the literature values listed here, $\Ebv$ is converted to $\Av$ using $R_\mathrm{V}=3.1$.}
\label{tab:gas-to-dust ratio}
\end{deluxetable*}

%% file: scale_height.tex
\startlongtable 
\begin{deluxetable*}{lccl}
    \tablecolumns{4}
    \tabletypesize{\scriptsize}
    \setlength{\tabcolsep}{0.1in}
    \tablewidth{0pt}
    \tablecaption{Comparisons of Scale Heights for Different Galactic Disk Components}
    \tablehead{
    \colhead{Reference} &
    \colhead{Scale Height} &
    \colhead{Component} &
    \colhead{Comment} \\
    \colhead{} &
    \colhead{(pc)} &
    \colhead{} &
    \colhead{} 
    }
    \startdata
This study & $43.3_{-3.5}^{+4.0}$ & MCs & MCs identified by excess reddening \\
\citet{2023ApJS..268....1D} & $\sim48$ & MCs & MCs traced by CO isotopologues and with $R\sim9$\,kpc \\
\citet{2014AA...564A.101L} & $\sim73$ & [C $\scriptstyle\mathrm{II}$] & [C $\scriptstyle\mathrm{II}$] traces a mix of ISM components \\
\citet{2006AA...446..121B} & $48 \pm 3$ & OCs & OCs with age $<200$\,Myr \\
\citet{2009RAA.....9.1285Z} & $51 \pm 5$ & OCs & OCs with age $\sim17$\,Myr \\
\citet{2016AA...593A.116J} & $60 \pm 2$ & OCs & OCs with age $<700$\,Myr and within 1.8\,kpc \\
\citet{2006AA...445..545P} & $56 \pm 3$ & OCs & OCs within 850\,pc from the Sun \\
\citet{2021AA...652A.102H} & $70.5 \pm 2.3$ & OCs & OCs with age $<20$\,Myr\\
\citet{2014MNRAS.444..290B} & $40-75$ & Clusters & Clusters with age from 1\,Myr to 1\,Gyr \\
\citet{2016BaltA..25..261B} & $46 \pm 5$ & Masers & Masers within 6\,kpc from the Sun\\
\citet{2016BaltA..25..261B} & $36 \pm 3$ & OB associations & OB associations within 3.5\,kpc from the Sun\\
\citet{2016BaltA..25..261B} & $35.6 \pm 2.7$ & \HII~regions & \HII~regions within 4.5 kpc from the Sun\\
\citet{2016BaltA..25..261B} & $52.1 \pm 1.9$ & Young Cepheids & Cepheids with age $\sim75$\,Myr and within 4\,kpc \\
\citet{2016BaltA..25..261B} & $72.0 \pm 2.3$ & Old Cepheids & Cepheids with age $\sim138$\,Myr and within 4\,kpc \\
\citet{2015ARAA..53..583H} & $\sim38-50$ & $\Htwo$ & $\Htwo$ traced by CO with $R$ from 2\,kpc to 8\,kpc\\
\citet{2017AA...607A.106M} & $64 \pm 12$ & $\Htwo$ & $\Htwo$ traced by CO inside the solar circle \\
\citet{2024ApJ...966..206W} & $ 61 \pm 9 $ & Cold \HI~clouds & \HI~absorbing clouds in the solar neighborhood\\
\citet{2021ApJ...906...47G} & $72.7 \pm 2.2$ & Dust & Thin dust disk \\
\citet{2018ApJ...858...75L} & $103.4^{+1.7}_{-1.8}$ & Dust & A single exponential disk \\
\citet{2017MNRAS.464.2545C} & $322 \pm 40$ & Stars & Stars from the outer disk and the halo \\
\enddata
\label{tab:scale_height}
\end{deluxetable*}